\documentclass[12 pt,twoside,aps,prd,amsmath,amssymb,
tightenlines,showpacs,showkeys]{revtex4}
\usepackage{mathptmx}
\usepackage{bm}
\usepackage{graphicx}
\usepackage[colorlinks]{hyperref}

\newcommand {\cG}{\cal G}
\newcommand {\cD}{\cal D}

\newcommand {\bg}{\bar \gamma}

\newcommand {\bp}{\bar \psi}

\def \myfigures #1#2#3#4#5#6#7#8
{\begin{figure}[ht]
    \begin{center}
        \includegraphics[width=#2 \textwidth]{#1.jpg}
        \hfill
        \includegraphics[width=#6 \textwidth]{#5.jpg}
        \parbox[t]{#4\textwidth}{\caption {#3}\label{#1}}
        \hfill
        \parbox[t]{#8\textwidth}{\caption {#7}\label{#5}}
    \end{center}
\end{figure} }

\def\myfigure #1#2#3#4
{\begin{figure}[ht]\begin{center}
\includegraphics[width=#2 \textwidth]{#1.eps}
\parbox[t]{#4\textwidth}{\caption{#3}\label{#1}}
\end{center}\end{figure}}

\begin{document}
\title{Spinors in cylindrically symmetric space-time}
\author{Bijan Saha}
\affiliation{Laboratory of Information Technologies\\
Joint Institute for Nuclear Research, Dubna\\
141980 Dubna, Moscow Region, Russia\\ and\\
Peoples' Friendship University of Russia (RUDN University)\\
6 Miklukho-Maklaya Street, Moscow, 117198, Russian Federation}
\email{bijan@jinr.ru} \homepage{http://spinor.bijansaha.ru}

\hskip 1 cm

\begin{abstract}

We studied the behavior of  nonlinear spinor field within the scope
of a static cylindrically symmetric space-time.  It is found that
the energy-momentum tensor (EMT) of the spinor field in this case
possesses nontrivial non-diagonal components. The presence of
non-diagonal components of the EMT imposes three-way restrictions
wither on the space-time geometry or on the components of the spinor
field or on both. It should be noted that analogical situation
occurs in cosmology when nonlinear spinor field is exploited as a
source of gravitational field given by Bianchi type-I cosmological
model.
\end{abstract}



\maketitle

\section{Introduction}

In recent years spinor field is being used in cosmology by many
authors \cite{SahaGRG1997,SahaPRD2001,Greene,Kremer,ELKO,Poplap}.
The ability of spinor field to simulate different kind of source
fields such as perfect fluid, dark energy etc.
\cite{SahaPRD2006,SahaECAA2014} allows one to study the evolution of
the Universe at different stages and consider the spinor field as an
alternative model of dark energy.

To our knowledge, except Friedmann-Robertson-Walker (FRW) model
given in Cartesian coordinates, in all other space-time spinor field
possesses nontrivial non-diagonal components of the energy-momentum
tensor. This very fact imposes severe restrictions on the geometry
of space-time and/or on the components of the spinor field
\cite{Saha2018ECAA}. As far as static spherically symmetric
space-time is concerned, the presence of non-diagonal components of
EMT imposes restrictions on the spinor field only \cite{BRS2020}.

Introduction of the spinor field in a classical theory such as
general relativity and cosmology gives rise to several questions due
to its quantum origin. Many specialists think that even if one uses
the spinor field in general relativity, he should treat it as
Grassmann variables. This is partially right, though we think that
spinors can be treated as classical complex projective coordinates
in the spirit of Dirac-Sommerfeld-Brioski
\cite{Mitskeevich,Cartan,Sommerfeld} as well. In this approach they
describe the condensation of "quark-antiquarks" and are ordinary
classical fields \cite{SahaAPSS2020}.

Note that spinor fields were introduced into the Einstein system
exploiting both quantum and classical interpretations.  A Fermi
field coupled to a homogeneous and isotropic gravitational field was
considered in \cite{Isham}, while the spinor was treated as a
Grassmann variable in \cite{SahaAPSS2015}. Dolan has studied the
Chiral Fermions and the torsion arising from it within the scope of
FRW geometries in the early Universe [\cite{Dolan}. In doing so he
argued that a quantum matter can be used as a source for the
classical field while the quantum aspects of the field itself can be
ignored.

As it was mentioned earlier, recently spinor field is being used in
astrophysics. Most of these works were done within the scope of
static spherically-symmetric space-time
\cite{BRS2020,SahaAPSS2020,Dzhun1}. Since a number of astrophysical
objects are given by cylindrically-symmetric space-time
\cite{broncqg2020} in this report we plan to consider the spinor
field within this model. String-like configurations of nonlinear
spinor field in a static cylindrically-symmetric space-time was
obtained in \cite{brongrg2003}. An interacting system of nonlinear
spinor and scalar fields in a static cylindrically-symmetric
space-time filled with barotropic gas was considered in
\cite{Shikin}. Unfortunately, the authors did not take into account
influence of the spinor field that occurs due to the presence of
non-diagonal components of EMT. In this paper I plan to address
those problems overlooked there and see if spinor field can be
exploited to construct different types of configurations seen in
astrophysics.

\section{Basic equations}

The action we choose in the form

\begin{eqnarray}
{\cal S} = \int \sqrt{-g} \left[\frac{R}{2 \kappa} + L_{\rm sp}
\right] d \Omega. \label{action}
\end{eqnarray}

where $\kappa = 8 \pi G$  is Einstein's gravitational constant, $R$
is the scalar curvature and $L_{\rm sp} $ is the spinor field
Lagrangian given by \cite{SahaPRD2001}

\begin{eqnarray}
L_{\rm sp} = \frac{\imath}{2} \biggl[\bp \gamma^{\mu} \nabla_{\mu}
\psi- \nabla_{\mu} \bar \psi \gamma^{\mu} \psi \biggr] - m \bp \psi
- \lambda F(K). \label{lspin}
\end{eqnarray}

To maintain the Lorentz invariance of the spinor field equations the
self-interaction (nonlinear term) $F(K)$  is constructed as some
arbitrary functions of invariants generated from the real bilinear
forms. On account of Fierz equality in \eqref{lspin} we set $K =
K(I, J) = b_1 I + b_2 J.$ Setting $(b_1=1, b_2=0),\, (b_1=0,
b_2=1),\, (b_1 = 1, b_2= 1),\,(b_1= 1, b_2= -1)$ for $K$ we obtain
one of the following expressions $\{I,\,J,\,I+J,\,I-J\}$. Here $I =
S^2$ and $J = P^2$ are the invariants of bilinear spinor forms with
$ S = \bp \psi$ and $P = \imath \bp \bg^5 \psi$ being the scalar and
pseudo-scalar, respectively. In \eqref{lspin} $\lambda$ is the
self-coupling constant.

The covariant derivatives of spinor field takes the form
\cite{SahaPRD2001}

\begin{eqnarray}
\nabla_\mu \psi = \partial_\mu \psi - \Omega_\mu \psi, \quad
\nabla_\mu \bp =
\partial_\mu \bp + \bp \Omega_\mu, \label{CVD}
\end{eqnarray}
where $\Omega_\mu$ is the spinor affine connection which can be
defined as \cite{SahaPRD2001}

\begin{eqnarray}
\Omega_\mu &=& \frac{1}{8}\left[\partial_\mu \gamma_\alpha,
\gamma^\alpha\right] - \frac{1}{8}
\Gamma^{\beta}_{\mu\alpha}\left[\gamma_\beta, \gamma^\alpha\right].
\label{SPAC}
\end{eqnarray}
Here $\left[a,b\right] = a b - b a$ and $\Gamma^{\beta}_{\mu\alpha}$
is the Christoffel symbol. In \eqref{SPAC} the Dirac matrices in
curve space-time $\gamma$ are connected to the flat space-time Dirac
matrices $\bg$ in the following way

\begin{eqnarray}
\gamma_\beta = e_\beta^{(b)} \bg_b, \quad \gamma^\alpha =
e^\alpha_{(a)} \bg^a, \nonumber
\end{eqnarray}
where $e^\alpha_{(a)}$ and $e_\beta^{(b)}$ are the tetrad vectors
such that

\begin{eqnarray}
e^\alpha_{(a)} e_\beta^{(a)} = \delta_\beta^\alpha, \quad
e^\alpha_{(a)} e_\alpha^{(b)} = \delta_a^b. \nonumber
\end{eqnarray}

The $\gamma$ matrices obey the following anti-commutation rules

\begin{eqnarray}
\gamma_\mu \gamma_\nu + \gamma_\nu \gamma_\mu = 2 g_{\mu\nu}, \quad
\gamma^\mu \gamma^\nu + \gamma^\nu \gamma^\mu = 2 g^{\mu\nu}.
\nonumber
\end{eqnarray}

Let us consider the cylindrically symmetric space-time given by
\begin{equation}
ds^2 = e^{2\gamma} dt^2 - e^{2 \alpha} du^2 - e^{2 \beta} d\phi^2 -
e^{2 \mu} dz^2,  \label{cylin}
\end{equation}

where $\gamma,\, \alpha,\, \beta$ and $\mu$ are the functions of the
radial coordinate $u$ only.

The tetrad we will choose in the form

\begin{align}
e^{(0)}_0 &= e^{\gamma}, \quad e^{(1)}_1 = e^{\alpha},\quad
e^{(2)}_2 = e^{\beta}, \quad e^{(3)}_3 = e^{\mu}, \nonumber\\
e^{0}_{(0)} &= e^{-\gamma}, \quad e^{1}_{(1)} = e^{-\alpha},\quad
e^{2}_{(2)} = e^{-\beta}, \quad e^{3}_{(3)} = e^{-\mu}. \nonumber
\end{align}

From $\gamma^\eta = e^\eta_{(a)} \bg^a$ we find

\begin{align}
\gamma^0 &= e^{-\gamma}\bg^0,\, \quad \gamma^1 = e^{-\alpha}
\bg^1,\, \quad \gamma^2 = e^{-\beta} \bg^2, \quad \gamma^3 =
e^{-\mu}\bg^3,
\nonumber\\
\gamma_0 &= e^{\gamma}\bg_0,\, \quad \gamma_1 = e^{\alpha} \bg_1,\,
\quad \gamma_2 = e^{\beta} \bg_2, \quad \gamma_3 = e^{\mu}\bg_3,
\nonumber
\end{align}
with
\begin{align}
\gamma_0 = \gamma^0,\quad \gamma_1 = -\gamma^1,\quad \gamma_2 =
-\gamma^2,\quad \gamma_3 = -\gamma^3. \nonumber
\end{align}

The nontrivial Christoffel symbols corresponding to the metric
\eqref{cylin} are
\begin{align}
\Gamma_{01}^{0} &= \gamma^\prime,\quad \Gamma_{11}^{1} =
\alpha^\prime,\quad
\Gamma_{21}^{2} = \beta^\prime, \quad \Gamma_{31}^{3} = \mu^\prime,  \nonumber\\
\Gamma_{00}^{1} &= e^{2(\gamma - \alpha)} \gamma^\prime,\quad
\Gamma_{22}^{1} = -e^{2(\beta -\alpha)} \beta^\prime, \quad
\Gamma_{33}^{1} = -e^{2(\mu -\alpha)} \mu^\prime. \nonumber
\end{align}

Then from the definition \eqref{SPAC} we find the following spinor
affine connections $\Omega_\mu$:

\begin{align}
\Omega_0 &= -\frac{1}{2} e^{\gamma - \alpha}\, \gamma^\prime\, \bg^0
\bg^1, \quad \Omega_1 = 0, \quad \Omega_2 = \frac{1}{2} e^{\beta -
\alpha}\, \beta^\prime\, \bg^2 \bg^1, \quad \Omega_3 = \frac{1}{2}
e^{\mu - \alpha}\, \mu^\prime\, \bg^3 \bg^1. \label{SAC}
\end{align}

The spinor field equations corresponding to the spinor field
Lagrangian \eqref{lspin} are \cite{SahaPRD2001}
\begin{subequations}
\label{speq}
\begin{align}
\imath\gamma^\mu \nabla_\mu \psi - m \psi - {\cD} \psi - \imath {\cG} \bg^5 \psi &=0, \label{speq1} \\
\imath \nabla_\mu \bp \gamma^\mu +  m \bp + {\cD} \bp  + \imath
{\cG} \bp \bg^5 &= 0, \label{speq2}
\end{align}
\end{subequations}

where we denote ${\cD} = 2 \lambda  F_K b_1 S, \quad {\cG} = 2
\lambda F_K b_2 P.$  On account of \eqref{speq} from \eqref{lspin}
one finds that $L_{\rm sp} = \lambda \left( 2 K F_K - F\right).$

Let the spinor field be a function of $u$ only, then in view of
\eqref{SAC} the spinor field equations can be written as

\begin{subequations}
\label{speqex}
\begin{eqnarray}
\psi^\prime + \frac{1}{2} \tau^\prime \psi - \imath e^{\alpha}
\left(m + {\cD}\right)
\bg^1 \psi - e^{\alpha}{\cG} \bg^5 \bg^1 \psi &=&0, \label{speq1n} \\
\bp^\prime + \frac{1}{2} \tau^\prime \bp + \imath e^{\alpha} \left(m
+ {\cD}\right) \bp \bg^1 - e^{\alpha}{\cG} \bp \bg^5 \bg^1  &=&0,
\label{speq2n}
\end{eqnarray}
\end{subequations}

where prime denotes differentiation with respect to $u$. In
\eqref{speqex} we also define
\begin{equation}
\tau = \left(\gamma + \beta + \mu\right). \label{taudef}
\end{equation}

The energy-momentum tensor of the spinor field is defined as
\cite{Pauli,brill,SahaPRD2001}

\begin{align}
T_{\mu}^{\,\,\,\rho}&=\frac{\imath}{4} g^{\rho\nu} \biggl(\bp
\gamma_\mu \nabla_\nu \psi + \bp \gamma_\nu \nabla_\mu \psi -
\nabla_\mu \bar \psi \gamma_\nu \psi - \nabla_\nu \bp \gamma_\mu
\psi \biggr) \,- \delta_{\mu}^{\rho} L_{\rm sp}  \nonumber\\
&= \frac{\imath}{4} g^{\rho\nu} \biggl(\bp \gamma_\mu
\partial_\nu \psi + \bp \gamma_\nu \partial_\mu \psi -
\partial_\mu \bar \psi \gamma_\nu \psi - \partial_\nu \bp \gamma_\mu
\psi \biggr) \nonumber \\ & - \frac{\imath}{4} g^{\rho\nu} \bp
\biggl(\gamma_\mu \Omega_\nu + \Omega_\nu \gamma_\mu + \gamma_\nu
\Omega_\mu + \Omega_\mu \gamma_\nu\biggr)\psi
 \,- \delta_{\mu}^{\rho} L_{\rm sp}. \label{temsp01}
\end{align}

From \eqref{temsp01} one finds the non-trivial components of the
energy-momentum tensor of the the spinor field
\begin{subequations}
\begin{align}
T_1^1 &=   m S + \lambda F, \label{11f} \\
T_0^0 &=  T_2^2 = T_3^ 3 = - \lambda \left( 2 K F_K -
F\right), \label{iif}\\
T^0_2 &= -\frac{\imath}{4} e^{\beta - \alpha - \gamma}\, \left(\gamma^\prime - \beta^\prime\right)\,A^3, \label{02f}\\
T^0_3 &= -\frac{\imath}{4} e^{\mu - \alpha - \gamma}\, \left(\gamma^\prime - \mu^\prime\right)\,A^2, \label{03f}\\
T^2_3 &= -\frac{\imath}{4} e^{\mu - \beta -
\alpha}\,\left(\beta^\prime - \mu^\prime\right)\,A^0, \label{23f}
\end{align}
\end{subequations}
with $A^\eta = \bp \bg^5 \bg^\eta \psi$ being the pseudovector. It
can be noticed that $T_0^0 + T_1^ 1 = m S + 2 \lambda \left(F - K
F_K\right)$ and $T_0^0 - T_1^ 1 =- \left( mS +  2 K F_K\right)$
might be positive or negative under certain conditions.

From \eqref{speqex} we find the following system of equations for
the bilinear spinor forms:

\begin{subequations}
\label{inveq}
\begin{align}
S^\prime + \tau^\prime S - 2 e^\alpha {\cG} A^1 &= 0, \label{S}\\
P^\prime + \tau^\prime P + 2 e^\alpha \left( m + {\cD}\right) A^1 &= 0, \label{P}\\
{A^1}^\prime +\tau^\prime A^1 + 2 e^\alpha \left( m + {\cD}\right) P
- 2 e^\alpha {\cG} S &= 0. \label{A1}
\end{align}
\end{subequations}
Equation \eqref{inveq} yields the following relation
\begin{equation}
S^2 + P^2 - \left(A^1\right)^2  = C_0 e^{-2 \tau}, \quad C_0 = {\rm
Const.} \label{inv}
\end{equation}

In case of $K = I$, i.e., ${\cal G} = 0$ from \eqref{S} we find

\begin{equation}
S = C_s e^{-\tau} \Rightarrow K = C_s^2 e^{-2\tau}. \label{Sx}
\end{equation}

If $K = J$, then in case of a massless spinor field from \eqref{P}
we find

\begin{equation}
P = C_p e^{-\tau} \Rightarrow K = C_p^2 e^{-2\tau}. \label{Sp}
\end{equation}

Let us consider the case when $K = I + J$. In this case $b_1 = b_2 =
1$. Then on account of expression for ${\cal D}$ and ${\cal G}$ from
\eqref{S} and \eqref{P} for the massless spinor field we find

\begin{subequations}
\begin{align}
S^\prime + \tau^\prime S - 4 e^\alpha \lambda F_K A^1 P  &= 0, \label{ipj1}\\
P^\prime + \tau^\prime P + 4 e^\alpha \lambda F_K A^1 S &= 0,
\label{ipj2}
\end{align}
\end{subequations}
which yields
\begin{equation}
K = I + J = S^2 + P^2 = C_1^2 e^{-2\tau}. \label{ipjf}
\end{equation}

Finally in case when $K = I - J$, i.e. $b_1 = - b_2 = 1$ from
\eqref{S} and \eqref{P} for the massless spinor field we obtain

\begin{subequations}
\begin{align}
S^\prime + \tau^\prime S + 4 e^\alpha \lambda F_K A^1 P  &= 0, \label{imj1}\\
P^\prime + \tau^\prime P + 4 e^\alpha \lambda F_K A^1 S &= 0,
\label{imj2}
\end{align}
\end{subequations}
which leads to
\begin{equation}
K = I - J = S^2 - P^2 = C_2^2e^{-2\tau}. \label{imjf}
\end{equation}

The Einstein tensor corresponding to the metric \eqref{cylin}
possesses only diagonal components. So let us first consider the
diagonal equations of Einstein system

\begin{subequations}
\begin{align}
e^{-2\alpha} \left[\gamma^\prime \beta^\prime + \beta^\prime
\mu^\prime + \mu^\prime \gamma^\prime\right] &= m S + \lambda F,
\label{EE11}\\
e^{-2\alpha} \left[\gamma^{\prime\prime} + \mu^{\prime\prime} +
\gamma^{\prime 2} + \mu^{\prime 2} + \gamma^{\prime} \mu^{\prime} -
\alpha^{\prime}\gamma^{\prime} - \alpha^{\prime}\mu^{\prime}\right]
&=   \lambda \left( F - 2 K F_K \right), \label{EE22}\\
e^{-2\alpha} \left[\gamma^{\prime\prime} + \beta^{\prime\prime} +
\gamma^{\prime 2} + \beta^{\prime 2} + \gamma^{\prime}
\beta^{\prime} - \alpha^{\prime}\gamma^{\prime} -
\alpha^{\prime}\beta^{\prime}\right] &=   \lambda \left( F - 2 K F_K
\right), \label{EE33}\\
e^{-2\alpha} \left[\beta^{\prime\prime} + \mu^{\prime\prime} +
\beta^{\prime 2} + \mu^{\prime 2} + \beta^{\prime} \mu^{\prime} -
\alpha^{\prime}\beta^{\prime} - \alpha^{\prime}\mu^{\prime}\right]
&=   \lambda \left(F - 2 K F_K\right). \label{EE00}
\end{align}
\end{subequations}

Subtraction of \eqref{EE00} from \eqref{EE22} yields

\begin{equation}
\gamma^{\prime\prime} - \beta^{\prime\prime} + \left(\gamma^{\prime}
- \beta^{\prime}\right)\left(\tau^\prime - \alpha^{\prime}\right) =
0, \label{2200}
\end{equation}
with the solution
\begin{equation}
\beta^\prime = \gamma^\prime - C_1 e^{\left(\alpha - \tau\right)},
\label{gambet}
\end{equation}
Analogically, subtracting \eqref{EE00} from \eqref{EE33} one finds
\begin{equation}
\mu^\prime = \gamma^\prime - C_2 e^{\left(\alpha - \tau\right)},
\label{gammu}
\end{equation}

In view of \eqref{taudef}, \eqref{gambet} and \eqref{gammu} one
finds

\begin{equation}
\gamma^\prime = \frac{1}{3} \left[\tau^\prime + \left(C_1 +
C_2\right) e^{\left(\alpha - \tau\right)}\right]. \label{gammatau}
\end{equation}

Thus, $\gamma$,\, $\beta$ and $\mu$ can be found in terms of
$\alpha$ and $\tau$. Let us find the equation for $\tau$. Summation
of \eqref{EE22}, \eqref{EE33}, \eqref{EE00} and 3 times \eqref{EE11}
gives

\begin{align}
e^{-2\alpha}\left[\tau^{\prime\prime} +  \tau^{\prime 2}  -
\alpha^\prime \tau^{\prime} \right] = \frac{3\kappa}{2} \left[mS + 2
\lambda \left(F - K F_K \right)\right]. \label{Eqtau}
\end{align}

Recall that for non-diagonal components of the EMT of the spinor
field we have non-trivial expressions, whereas the non-diagonal
components of the Einstein tensor in this case are trivial. Equating
these expressions to zero from \eqref{02f}, \eqref{03f} and
\eqref{23f} we obtain the following constrains

\begin{subequations}
\begin{align}
\left(\gamma^\prime - \beta^\prime\right)\,A^3 &= 0, \label{02c}\\
\left(\gamma^\prime - \mu^\prime\right)\,A^2 &= 0, \label{03c}\\
\left(\beta^\prime - \mu^\prime\right)\,A^0 &= 0. \label{23c}
\end{align}
\end{subequations}

The foregoing expressions give rise to three possibilities:

\begin{subequations}
\begin{align}
A^3 &= A^2 = A^0 = 0\quad {\rm and} \quad \gamma^\prime \ne
\beta^\prime \ne \mu^\prime \Rightarrow C_1 \ne C_2 \ne 0 \label{pos1}\\
A^2 &= A^0 = 0 \quad {\rm and} \quad \gamma^\prime - \beta^\prime =
0 \Rightarrow C_1 = 0,  \label{pos2}\\
A^3,&\,\, A^2,\,\, A^0 \quad {\rm are\,\, nontrivial\,\, and} \quad
\gamma^\prime = \beta^\prime = \mu^\prime \Rightarrow C_1 = C_2 = 0.
\label{pos3}
\end{align}
\end{subequations}
It should be noted that in a Bianchi type-I space-time there occur
similar possibilities \cite{SahaAPSS2015}. In that case under the
assumption \eqref{pos1} the spinor field becomes massless and linear
\cite{SahaAPSS2015}. In a static cylindrically symmetric space-time
that is not necessarily the case.

Unfortunately, right now we cannot exactly solve the equation for
defining either $\tau$ or $\alpha$. So we have to assume some
coordinate conditions. There might be a few. In what follows, we
consider the case with $K = I$, as in this case it is possible to
consider massive spinor. Further we set $S = K_0 e^{-\tau}$ and $K =
K_0^2 e^{-2\tau}.$

{\bf Case 1} Let us first consider the harmonic radial coordinate
$u$ such that the following relation holds for the metric functions
\cite{bron}:

\begin{equation}
\alpha = \gamma + \beta + \mu. \label{hc}
\end{equation}

In view of \eqref{taudef} Eq. \eqref{Eqtau} takes the form

\begin{align}
\tau^{\prime\prime}  = \frac{3\kappa}{2}  e^{2\alpha} \left[mS + 2
\lambda \left(F - K F_K \right)\right]. \label{Eqtau1}
\end{align}

Let us consider the case when $F$ is a power law function of $K$,
i.e. $F = K^{n+1}$. Inserting $S = K_0 e^{-\tau}$ and $K = K_0^2
e^{-2\tau}$ into \eqref{Eqtau} on account of $\alpha = \tau$ we find

\begin{align}
\tau^{\prime\prime}  = \frac{3\kappa}{2} m K_0 e^{\tau} - 3 \kappa
\lambda n K_0^{2(n+1)}e^{-2 n \tau}, \label{Eqtau2}
\end{align}
with the first integral
\begin{equation}
\tau^\prime = \sqrt{3\kappa m K_0 e^{\tau} + 3 \kappa \lambda
K_0^{2(n+1)}e^{-2 n \tau} + C_3}, \quad C_3 = {\rm const.}
\label{1stint0}
\end{equation}
So the solution can be given in quadrature

\begin{equation}
\int \frac{d\tau}{\sqrt{3\kappa m K_0 e^{\tau} + 3 \kappa \lambda
K_0^{2(n+1)}e^{-2 n \tau} + C_3}} = u + u_0, \quad u_0 = {\rm
const.} \label{int}
\end{equation}

Let us consider some simple cases those allow exact solution.

Fisrt we study the Heisenberg-Ivanenko type nonlinearity when $F(K)
= K$. It can be obtained by setting $n = 0$ in \eqref{1stint0}. In
this case \eqref{1stint0} takes the form

\begin{equation}
\tau^\prime = \sqrt{3 \kappa m K_0 e^{\tau} + C_4}, \quad C_4 = C_3
+ 3 \kappa \lambda K_0^2, \label{1st}
\end{equation}

which finally gives

\begin{equation}
e^{\tau} = e^{\alpha} =\frac{C_4}{2\kappa m K_0\,
\sinh^2{\left(-\sqrt{C_4}u/2 + C_5\right)}}, \quad C_5 = {\rm
const.} \label{soltau0}
\end{equation}

For a general power law type nonlinearity we study the massless
spinor field. Setting $m = 0$ in \eqref{1stint0} we have

\begin{equation}
\tau^\prime = \sqrt{3\kappa K_0^{2(n+1)} e^{-2n\tau} + C_3},
\label{1st2}
\end{equation}
with the solution
\begin{equation}
e^\tau = e^{\alpha} = \left[\sqrt{\frac{3\kappa K_0^{2(n+1)}}{C_3}}
\sinh{\left(n \sqrt{C_3}\, u + C_6\right)}\right]^{1/n}, \quad C_6 =
{\rm const.} \label{fin1}
\end{equation}

For a more general solution to the Einstein equations with massive
and nonlinear spinor field as source we rewrite it in the form of
Cauchy:
\begin{subequations}
\begin{align}
\tau^\prime &= \eta, \label{eta}\\
\eta^\prime &= \frac{3\kappa}{2} e^{2\tau} \left[mS + 2\left(F - K
F_K \right)\right],
\label{taueta}\\
\gamma^\prime &= \frac{1}{3} \left[\eta + \left(C_1 +
C_2\right)\right], \label{gam3}\\
\beta^\prime &= \frac{1}{3}\left(\eta - 2 C_1 + C_2\right),\label{bet3}\\
\mu^\prime &= \frac{1}{3}\left(\eta + C_1 - 2 C_2\right).\label{mu3}
\end{align}
\end{subequations}

This system can be solved numerically. In Figures 1 and 2, we have
plotted the metric functions $\gamma(u), \alpha(u), \beta(u),
\mu(u)$ for different types of nonlinearities, namely, $n=0$
(Heisenberg-Ivanenko case) and $n =4$. For simplisity, we set the
following values for other parameters $C_1 = 1$, $C_2 = 2$ and $m =
1$. The initial values were taken to be $\tau (0) =0.3$, $\gamma (0)
= 0.03$, $\beta (0) = 0.2$, $\mu (0) = 0.07$ and $\nu(0) = 0.2$. As
we see from the graphics, with the increase of the value of $n$ the
difference between the metric functions increases.

\vskip 1 cm \myfigures{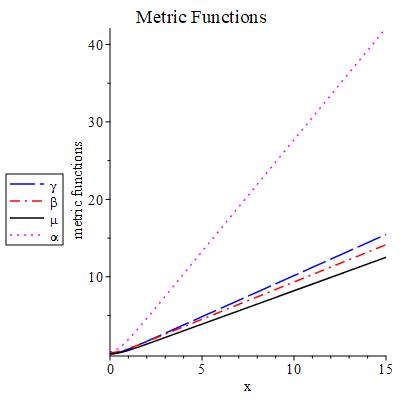}{0.46}{Plot of metric
functions for Heisenberg--Ivanenko type nonlinearity for harmonic
radial coordinate.}{0.45}{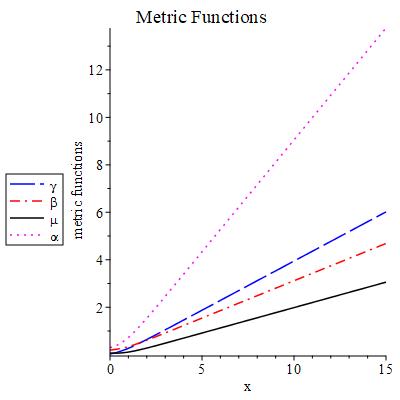}{0.43}{Plot of metric
functions for a massive nonlinear spinor field with power on
nonlinearity $n = 4$ for harmonic radial coordinate.}{0.45}

{\bf Case 2} Let us consider the quasiblogal coordinate $\alpha = -
\gamma$ \cite{BronBook}. In this case for $\tau$ we have

\begin{align}
\tau^{\prime\prime} +  \tau^{\prime 2}  + \gamma^\prime
\tau^{\prime}  = \frac{3\kappa}{2} e^{-2\gamma} \left[mS + 2\left(F
- K F_K \right)\right], \label{Eqtauag}
\end{align}

whereas inserting \eqref{gambet} and \eqref{gammu} into \eqref{EE11}
for $\gamma$ we find

\begin{eqnarray}
3 \gamma^{\prime 2} - 2\left(C_1 + C_2\right) e^{-\left(\gamma +
\tau\right)}\, \gamma^\prime + C_1 C_2 e^{-2 \left(\gamma +
\tau\right)}  = e^{- 2 \gamma} \left(m S + \lambda F\right).
\label{gam2}
\end{eqnarray}

Let us rewrite \eqref{Eqtauag} and \eqref{gam2} in the Cauchy form

\begin{subequations}
\begin{align}
\tau^\prime &= \eta, \label{eta1}\\
\eta^\prime &= - \eta^2 -  \frac{\eta}{3} \left(C_1 + C_2\right)
e^{-\left(\gamma + \tau\right)} \mp  \frac{D}{6} \eta  +
\frac{3\kappa}{2} e^{-2\gamma} \left[mS + 2\left(F - K F_K
\right)\right],
\label{taueta1}\\
\gamma^\prime &= \frac{1}{3} \left(C_1 + C_2\right) e^{-\left(\gamma
+ \tau\right)} \pm \frac{D}{6}, \label{gam4}\\
\beta^\prime &= \frac{1}{3}\left(-2 C_1 + C_2\right)
e^{-\left(\gamma + \tau\right)} \pm \frac{D}{6},\label{bet4}\\
\mu^\prime &=  \frac{1}{3}\left(C_1 - 2 C_2\right) e^{-\left(\gamma
+ \tau\right)} \pm \frac{D}{6},\label{mu4}
\end{align}
\end{subequations}
where $D = 2\sqrt{\left(C_1 + C_2\right)^2 e^{-2\left(\gamma +
\tau\right)} - 3\left( C_1 C_2 e^{-2 \left(\gamma + \tau\right)} -
e^{- 2 \gamma} \left(m S + \lambda F\right)\right)}.$ As one sees,
the above-going system is valid if and only if $D \ge 0$. The
equation \eqref{gam4} is found from \eqref{gam2}, which is a
quadratic equation with respect to $\gamma^\prime$.

In the Figures 3 and 4, we have plotted the metric functions for the
same values as in previous cases, i.e. we set $C_1 = 1$, $C_2 = 2$
and $m = 1$ and the initial values were taken to be $\tau (0) =0.3$,
$\gamma (0) = 0.03$, $\beta (0) = 0.2$, $\mu (0) = 0.07$ and $\nu(0)
= 0.2$. Here we have consider the cases with $n = 0$ and $n = 4$.
And like the previous cases we see with the increase of $n$ the
difference between the metric functions increases.

\vskip 1 cm \myfigures{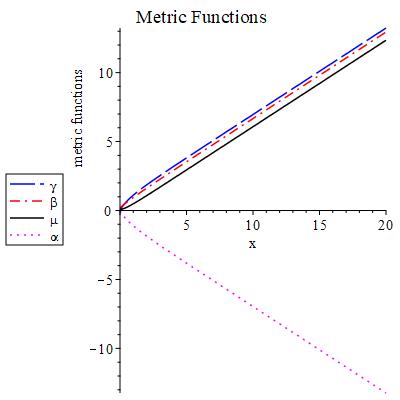}{0.46}{Plot of metric
functions for Heisenberg--Ivanenko typenonlinearity with $n = 0$ for
quasiglobal coordinate.}{0.45}{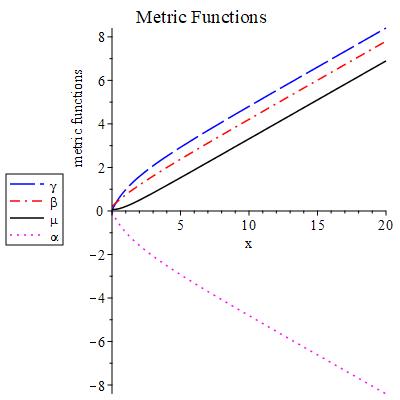}{0.43}{Plot of metric
functions for a nonlinear spinor field with power on nonlinearity $n
= 4$ with nonzero mass.}{0.45}

\section{Conclusions}

We studied a system of nonlinear spinor field minimally coupled to a
static cylindrically symmetric space-time. It is found that the
energy-momentum tensor (EMT) of the spinor field has nontrivial
non-diagonal components. The presence of non-diagonal components of
the EMT imposes three-way restrictions on the space-time geometry
and/or on the components of the spinor field. It should be noted
that such situation occurs in Bianchi type-I cosmological model as
well, but while in BI model under a specific type of restriction the
spinor field becomes massless and linear, this is not the case in
this model. Moreover, while in a static spherically symmetric
space-time the presence of non-trivial non-diagonal components of
EMT of the spinor field has no effect on the space-time geometry, in
static cylindrically symmetric space-time it influences both the
space-time geometry and the components of the spinor field. In the
present model we have $T_0^0 = T_2^2$ which in our view may play
crucial role in the formation of configurations like wormhole. It
should be noted that the expressions $(T_0^0 + T_1^1)$ and $(T_0^0 -
T_1^1)$ can be both positive and negative, depending on the type of
nonlinearity. In our view this fact may provide some very
interesting results which we plan to study in our upcoming papers.

\vskip 7mm

\noindent {\bf Acknowledgments}\\
The publication was prepared with the support of the "RUDN
University Program 5-100" and also partly supported by a joint
Romanian-JINR, Dubna Research Project, Order no.396/27.05.2019 p-71.

\end{document}